\documentclass[runningheads]{llncs}
\usepackage[T1]{fontenc}
\usepackage{graphicx}
\usepackage{booktabs}
\usepackage{amsmath}
\usepackage{tikz}
\usetikzlibrary{arrows.meta,positioning,calc}
\usepackage{algorithm}
\usepackage{algpseudocode}

\usepackage{array}
\usepackage{tabularx}
\newcolumntype{Y}{>{\raggedright\arraybackslash}X}
   
\newcommand{\ASR}{\mathrm{ASR}}
\newcommand{\score}{S}
\newcommand{\hashfun}{\mathcal{H}}
\newcommand{\dist}{d}
\newcommand{\lpips}{\mathrm{LPIPS}}
\newcommand{\queries}{Q}

\begin{document}
\title{LLM-Guided Program Evolution for Targeted
Black-Box Attacks on Perceptual Hash Algorithms}
\titlerunning{LLM-Guided Program Evolution for Targeted
Black-Box Attacks on PHAs}
%
\author{Aleksei S. Krylov\inst{1,2}\orcidID{0009-0009-7990-6895} \and
Denis S. Rakhov\inst{2}\orcidID{0009-0005-2915-7225} \and Veronica Veselova\inst{3}\orcidID{0009-0003-9508-0698} \and Dmitry Bolokhov\inst{3}\orcidID{1111-2222-3333-4444} \and 
Oleg Y. Rogov\inst{1,3,4,5}\orcidID{2222--3333-4444-5555}}

\authorrunning{A. Krylov et al.}


%
\institute{MIPT \and
Sberbank \and Central University \and
AIRI \and MTUCI-Labs}

%
\maketitle              
\begin{abstract}
Perceptual hash algorithms (PHAs) are widely deployed
to detect image forgery under benign transformations, yet their robustness against adversarially chosen perturbations remains poorly
understood and rarely comes with provable guarantees. We propose
a novel evolutionary framework based on GigaEvo and OpenEvolve
for targeted second-image attacks on perceptual hash algorithms.
We assess attack performance using a composite score that jointly accounts for the fraction of adversarial images whose normalized Hamming distance to the target hash falls below threshold p (Attack Success Rate), the number of queries issued to the hash function, and the L2 distortion relative to the original image. Experiments on four deployed PHAs (pHash, PDQ, PhotoDNA, NeuralHash) across 30 ImageNet image pairs demonstrate that our evolutionary approach achieves comparable or better ASR than existing black-box baselines using substantially fewer queries to the hash function, while simultaneously producing adversarial images with lower L2 distortion relative to the originals. The best evolved programs reduce the pre-defined composite attack score relative to the best optimized seed by 41.2\% for NeuralHash, 38.3\% for PDQ, 34.0\% for pHash, and 8.1\%  for PhotoDNA. Unlike gradient-based
methods, our framework requires no internal knowledge of PHA architectures and naturally handles the non-differentiable, discretized
nature of hash outputs. These results reveal previously unreported
vulnerabilities in widely deployed content-moderation pipelines and
motivate the development of provably robust perceptual hashing
schemes.


\end{abstract}
\section{Introduction}
Perceptual hash algorithms (PHAs) map multimedia objects to short binary strings so that
perceptually similar inputs yield nearby hashes under Hamming distance. This
functionality makes PHAs attractive for large-scale near-duplicate search \cite{10.1007/978-3-642-03079-6_15}, copyright enforcement,
and safety applications such as client- or server-side detection of known harmful imagery
\cite{farid2021overview}. In contrast to cryptographic hashing-where a single-bit change should
avalanche-perceptual hashing is intentionally stable under benign content-preserving
transformations (e.g., resizing, compression, mild blur), while remaining discriminative across distinct content. Popular deployed designs include DCT-based global hashes (pHash, PDQ)
\cite{klinger2013phash,davis2019open}, gradient aggregation schemes (PhotoDNA-like)
\cite{ith2015microsoft,burgett2014photodna,arthur2013twitter}, and neural embedding approaches
(NeuralHash-like) \cite{cobbe2021data}.
This stability--discrimination tension is not merely an engineering trade-off: it is a mathematical
constraint that becomes security-critical when an adversary can choose inputs. A substantial
empirical literature demonstrates that many PHAs are robust to common non-adversarial edits \cite{Zauner2010ImplementationAB,9850288}. However, recent work has also established that
robustness under benign transformations does not imply robustness against adversarial perturbations, including gradient-based hash-evasion attacks in white-box settings \cite{Dolhansky2020AdversarialCA,Struppek_2022} and practical attacks spanning inversion and evasion regimes \cite{madden2024robustnesspracticalperceptualhashing}. Moreover, for certain semantic classes (notably faces), the effective hash space can be far smaller than the nominal bit-length would suggest \cite{cryptoeprint:2024/1869}. These results raise a foundational
question: what can we prove about the robustness of a perceptual hash, and what instance-wise guarantees can a deployed system offer?
The central obstacle is that PHAs combine continuous feature extraction with discrete quantization.
Quantization is indispensable for fast indexing and compact storage, but it turns small feature
changes into potentially large Hamming changes whenever a coordinate crosses a threshold. While
this viewpoint is implicit in engineering descriptions of pHash/PDQ and related schemes
\cite{klinger2013phash,davis2019open}, it has not been systematically developed into a sharp,
per-instance robustness theory.
In this paper we provide a systematic empirical study of targeted second-image attacks on perceptual hash algorithms through the lens of evolutionary computation. Concretely, we make the following contributions. First, we propose a black-box evolutionary attack framework built on GigaEvo\cite{khrulkov2025gigaevoopensourceoptimization} and OpenEvolve\cite{novikov2025alphaevolvecodingagentscientific,openevolve} that searches for adversarial perturbations of a source image such that the resulting hash closely matches that of a designated target image. Second, we evaluate the framework against four deployed PHAs — pHash, PDQ, PhotoDNA, and NeuralHash — on 30 ImageNet image pairs, demonstrating that our approach consistently achieves higher ASR than existing black-box baselines while producing adversarial images with lower L2 distortion relative to their originals. Third, we show that evolutionary search is particularly well-suited to this problem: it requires no gradient information, no knowledge of internal PHA parameters, and naturally accommodates the non-differentiable, discrete structure of hash outputs. Together, these results deepen our understanding of the practical attack surface of deployed PHAs and highlight the need for robustness guarantees that account for adversarially chosen inputs.
\section{Related Work}
\label{sec:related}

\subsection{Perceptual Hash Algorithms}
Perceptual Hash Algorithms, shortly PHAs, have emerged as powerful technology to limit the redistribution of multimedia content (e.g. images) \cite{farid2021overview}. We consider the most popular of them
(Table \ref{tab:hashing_algorithms_comparison}):





\begin{table}
\caption{Comparison of perceptual hashing algorithms used in instant messaging platforms.}\label{tab:hashing_algorithms_comparison}
\begin{tabular}{|l|l|l|l|l|l|}
\hline
Algorithm & Developer & Release & Scale & Feature Extraction & Hash Length\\
\hline
\textbf{pHash} & Open Source & 2008 & Global & 2D DCT & 64 bits\\
\textbf{PDQ} & Meta & 2019 & Global & 2D DCT & 256 bits\\
\textbf{PhotoDNA} & Microsoft & 2009 & Mid-scale & $6\times 6$ Grid Gradients & 1152 bits\\
\textbf{NeuralHash} & Apple & 2021 & Pixel-scale & Deep CNN & 96 bits\\
\hline
\end{tabular}
\end{table}

\subsection{Robustness of the Perceptual Hash Algorithms}
Previous research has challenged the robustness of these perceptual hashing algorithms from various perspectives. The first of these studies proved that PHAs are resistant to resizing, recoloring, cropping, and blurring transformations \cite{9850288,Zauner2010ImplementationAB}. Nevertheless, vulnerabilities to adversarial attack were detected. Initially, researchers construct white-box gradient-based approaches \cite{Dolhansky2020AdversarialCA,Struppek_2022}. They have shown that although perceptual hashing-based detection systems may be cryptographically well-proven, they remain susceptible to complex collision attacks, which pose significant risks to user privacy and could facilitate surveillance far beyond the intended scope of CSAM detection. Furthermore, another study has shown that for certain image categories like faces, the effective hash space of NeuralHash (96-bit) is constrained, making random collisions likely in sets of only $2^{16}$ images \cite{cryptoeprint:2024/1869}. 
Adversarial attacks comprise two main groups: hash-inversion attacks and hash-evasion attacks. Hash-inversion attacks aim at inverting the hash function to obtain the original image from the hash \cite{madden2024robustnesspracticalperceptualhashing}. Hash-evasion attacks may be targeted and untargeted. In case of targeted attacks, researchers aim to obtain the hash of the target image for the current image utilizing visually imperceptible changes \cite{285409}. During an untargeted attack, it is enough to get a large difference in hashes with a small difference in images. To improve the quality of attacks, researchers also utilize multiresolution perturbation, where each perturbation element can affect image regions of adjustable scales \cite{307842}. Authors of this paper introduce three levels of perturbations: Global-scale, Mid-scale and Pixel-scale. Moreover, Atkscopes \cite{307842} demonstrate a substantial improvement in efficiency, compared to
previous attacks with mismatched scales by Prokos \cite{285409}.

\subsection{LLM--based evolution}
Evolutionary computation has a well--established tradition in combinatorial optimization. Its integration with large language models represents an emerging and rapidly advancing paradigm. One of the first significant papers in this domain was the AlphaEvolve paper \cite{novikov2025alphaevolvecodingagentscientific}. The authors achieved state--of--the--art results in the code generation task. After that, OpenEvolve framework was presented as an open source implementation of Google DeepMind's AlphaEvolve \cite{openevolve}. And other researchers have presented GigaEvo, an open-source framework for LLM--driven evolutionary computation with modular implementations of key components: MAP--Elites quality diversity algorithms, asynchronous DAG--based evaluation pipelines, LLM mutation operators with
bidirectional lineage tracking, and flexible configuration management - enabling rapid prototyping and improving upon prior results \cite{khrulkov2025gigaevoopensourceoptimization}. Beyond code generation and optimization, LLM-guided evolutionary search has also been applied to the automated design of methods, for example in uncertainty quantification \cite{seleznyov-etal-2026-evolutionary}.
To the best of our knowledge, no prior work has applied LLM-driven quality-diversity evolution to adversarial attacks on perceptual hashing algorithms. The present work introduces this approach in the context of targeted second-preimage collision attacks.

\section{Method}
\label{sec:method}

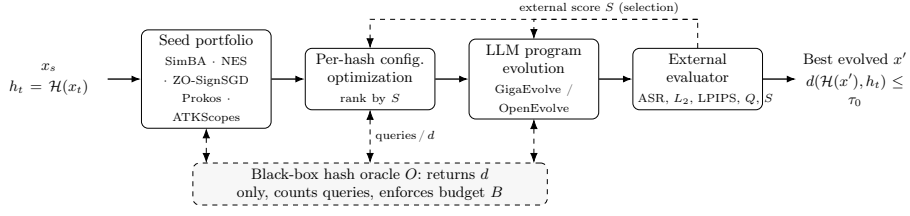
\begin{figure}[t]
\centering
\resizebox{\textwidth}{!}{%
\begin{tikzpicture}[
  font=\footnotesize,
  >={Latex[length=2mm]},
  node distance=7mm and 7mm,
  block/.style={draw, rounded corners, align=center, inner sep=3pt,
                minimum height=12mm, text width=25mm},
  io/.style={align=center, inner sep=2pt, text width=23mm},
  oracle/.style={draw, dashed, rounded corners, align=center, inner sep=3pt,
                 text width=72mm, fill=black!3},
  flow/.style={->, thick},
  q/.style={<->, dashed},
  fb/.style={->, dashed}
]
\node[io] (in) {$x_s$\\[2pt]$h_t=\hashfun(x_t)$};
\node[block, right=of in]   (seed) {Seed portfolio\\[1pt]{\scriptsize SimBA $\cdot$ NES $\cdot$ ZO-SignSGD}\\{\scriptsize Prokos $\cdot$ ATKScopes}};
\node[block, right=of seed]  (opt) {Per-hash config.\ optimization\\[1pt]{\scriptsize rank by $\score$}};
\node[block, right=of opt]   (evo) {LLM program evolution\\[1pt]{\scriptsize GigaEvolve / OpenEvolve}};
\node[block, right=of evo]  (eval) {External evaluator\\[1pt]{\scriptsize $\ASR,\,L_2,\,\lpips,\,\queries,\,\score$}};
\node[io, right=of eval]     (out) {Best evolved $x'$\\[2pt]$\dist(\hashfun(x'),h_t)\le\tau_0$};

\draw[flow] (in)   -- (seed);
\draw[flow] (seed) -- (opt);
\draw[flow] (opt)  -- (evo);
\draw[flow] (evo)  -- (eval);
\draw[flow] (eval) -- (out);

\node[oracle, below=11mm of opt] (oracle)
  {Black-box hash oracle $O$: returns $\dist$ only, counts queries, enforces budget $B$};
\draw[q] (seed.south) -- (seed.south |- oracle.north);
\draw[q] (opt.south)  -- (oracle.north)
         node[midway, right, font=\scriptsize]{queries\,/\,$\dist$};
\draw[q] (evo.south)  -- (evo.south |- oracle.north);

\coordinate (r) at ($(eval.north)+(0,6mm)$);
\draw[dashed] (eval.north) -- (r);
\draw[fb] (r) -| (evo.north);
\draw[fb] (r) -| (opt.north)
         node[pos=0.15, above, font=\scriptsize]{external score $\score$ (selection)};
\end{tikzpicture}}

\caption{Proposed framework. Per-hash tuned seed attacks seed LLM-guided program evolution against the black-box oracle $O$ returning only the hash distance $\dist$. A single external evaluator scores each candidate with $\score$ (lower is better).}
\label{fig:pipeline}
\end{figure}

\textbf{Targeted Second-Image Attack Formulation.} Let $x_s$ be a source image, $x_t$ a target image, and $h_t=\hashfun(x_t)$ the target hash. A targeted second-image attack searches for an adversarial image $x'$ close to $x_s$ such that
\begin{equation}
    \dist(\hashfun(x'), h_t) \leq \tau_0,
    \label{eq:success}
\end{equation}
where $\dist$ is Hamming distance for binary hashes and $L_1$ distance for PhotoDNA descriptors. The target image is used by the benchmark only to compute $h_t$. Candidate attacks receive the source image, the target hash through an oracle, the matching threshold, and query feedback; they do not receive gradients, internal hash features, or the target image itself.

\textbf{Matching Thresholds.} The matching threshold $\tau_0$ is fixed per hash and is not optimized by the attack. We use the operating points summarized in Table~\ref{tab:hash_algorithms}: 12 for pHash, 92 for PDQ, 3855 for PhotoDNA, and 17 for NeuralHash. A run is counted as successful only when the externally recomputed distance is within the corresponding threshold.

\begin{table}[t]
\centering
\caption{Perceptual hashing algorithms considered in the study.}
\label{tab:hash_algorithms}
\small
\begin{tabularx}{\textwidth}{@{}lYccY@{}}
\toprule
Hash & Descriptor & Distance & Threshold $\tau_0$ & Dominant feature scale \\
\midrule
pHash & 64-bit binary & Hamming & 12 & Global DCT structure \\
PDQ & 256-bit binary & Hamming & 92 & Global DCT structure \\
PhotoDNA & 144-dimensional numeric descriptor (1152 bits) & $L_1$ & 3855 & Mid-scale gradient structure \\
NeuralHash & 96-bit binary & Hamming & 17 & Neural image embedding \\
\bottomrule
\end{tabularx}
\end{table}

\textbf{Hash Similarity Loss.} For a set of $N$ image pairs, attack success rate is
\begin{equation}
    \ASR =
    \frac{1}{N}
    \sum_{i=1}^{N}
    \mathbf{1}\left[
        \min_j \dist(\hashfun(x'_{i,j}), h_{t,i}) \leq \tau_0
    \right],
    \label{eq:asr}
\end{equation}
where $j$ indexes candidates observed during the attack. The oracle tracks the best hash-distance candidate during a run, and attacks return an $x_\mathrm{best}$ image for external evaluation. Using the best observed candidate avoids penalizing an attack whose final iterate is worse than an earlier candidate.

\textbf{Visual Distortion Constraint.} Visual cost is measured by pixel-space $L_2$ and LPIPS \cite{zhang2018unreasonableeffectivenessdeepfeatures}. Query cost is the number of hash-oracle calls. The main ranking criterion is the implementation score
\begin{equation}
    \score =
    \frac{
        \overline{L_2}
        + 0.02\,\overline{\queries}
    }{
        \max(\ASR, 0.05)
    }.
    \label{eq:score}
\end{equation}
In the implementation, a technical zero-success penalty is added:
\begin{equation}
    \score \leftarrow \score + 10000 \cdot \mathbf{1}[\ASR=0].
\end{equation}
Lower score is better. The means in Eq.~\ref{eq:score} are computed over all evaluated pairs, including failures. The denominator penalizes low-ASR attacks; among successful attacks, the numerator favors lower $L_2$ distortion and fewer queries. The coefficient $0.02=1/50$ gives 50 extra oracle calls the same score cost as one unit of $L_2$. The zero-success penalty prevents attacks with no collisions from being ranked as plausible low-distortion candidates. LPIPS is reported separately and is not part of $S$ in the current benchmark tables, so a lower score should not be read as perceptual closeness when LPIPS remains high.

\begin{table}[t]
\centering
\caption{Metrics reported by the external evaluator.}
\label{tab:metrics}
\footnotesize
\begin{tabularx}{\textwidth}{@{}lYc@{}}
\toprule
Metric & Meaning & Direction \\
\midrule
ASR & Fraction of image pairs for which the returned candidate reaches the target matching threshold. & Higher \\
$L_2$ & Euclidean distance between the source image and the returned candidate in RGB pixel space. & Lower \\
LPIPS & Learned perceptual image patch similarity between the source image and the returned candidate. & Lower \\
Queries & Number of hash-oracle calls spent by the attack; the initial source-to-target distance is not counted. & Lower \\
Final hash distance & Distance between the returned candidate hash and the target hash after external recomputation. & Lower \\
$S$ & Composite score used for configuration optimization and program evolution. & Lower \\
\bottomrule
\end{tabularx}
\end{table}

\textbf{Optimization via Evolutionary Search.} All attacks are executed through a common functional interface,
\begin{equation}
    \texttt{run\_attack}(x_s, O, \theta, B, r) \rightarrow R,
    \label{eq:attack_contract}
\end{equation}
where $O$ is the hash oracle, $\theta$ are attack parameters, $B$ is the query budget, $r$ is a random seed, and $R$ contains the best image found by the attack. This deliberately simple contract is used for both hand-written baselines and LLM-generated Python programs.

The oracle is the only source of feedback. It stores the target hash, clips candidate images to the valid pixel range, counts queries, enforces the budget, and returns only the current distance to the target hash. This is a graded-distance black-box oracle, not a hard-label match/no-match API; hard-label and rate-limited settings are outside the present evaluation. The initial distance from $x_s$ to $h_t$ is computed when the oracle is created but is not counted as an attack query. During the run the oracle maintains a best-so-far candidate by hash distance.

The evaluator is the final source of truth. After an attack returns, it recomputes $\hashfun(x_\mathrm{best})$, success, final hash distance, $L_2$, LPIPS when requested, query count, and score. It does not trust success flags or metrics reported by the attack itself. This is especially important for program evolution: generated code can search for better candidates, but it cannot improve the reported result by changing bookkeeping or output formatting. The same evaluator is used for original baselines, optimized configurations, and evolved programs.

\noindent\textbf{Configuration optimization.}
Configuration optimization is performed independently for each hash using the Optuna framework \cite{akiba2019optunanextgenerationhyperparameteroptimization} with a tree-structured Parzen estimator sampler \cite{watanabe2025treestructuredparzenestimatorunderstanding}. The first phase searches for configurations that reach the threshold reliably; the second phase ranks configurations by the score in Eq.~\ref{eq:score}. This two-phase design avoids selecting visually clean attacks that rarely collide, while still preferring lower distortion and fewer queries once success is achieved. It also makes the comparison fair across hashes with different descriptor lengths, distance distributions, and useful perturbation scales. The resulting optimized configurations form the seed context for the LLM-evolution stage.

\noindent\textbf{Program evolution.}
For each hash, GigaEvolve and OpenEvolve receive the optimized attack implementations, empirical notes, and the evaluator contract. Candidate programs are generated by LLM mutation and executed on the benchmark image pairs. A valid candidate must implement the interface in Eq.~\ref{eq:attack_contract}; invalid programs, runtime failures, and malformed results are assigned penalty scores by the evaluator. OpenEvolve maximizes a framework-native score, so the adapter exposes $-S$ while preserving all original metrics for analysis. GigaEvolve uses the same evaluator and stores candidate lineage during search.

The evolved changes are algorithmic rather than cosmetic: the search basis, update schedule, restarts, clipping strategy, plateau handling, and post-success refinement can all be modified by the generated code, but the scoring function and benchmark pairs remain external to the candidate program.

\begin{table}[t]
\centering
\caption{Integration of the two program-evolution backends.}
\label{tab:evolution_backends}
\footnotesize
\begin{tabularx}{\textwidth}{@{}lYY@{}}
\toprule
Property & GigaEvolve & OpenEvolve \\
\midrule
Seed context & Optimized attack implementations, cleaned public parameters, empirical hash-specific notes, and the same evaluator contract. & The same seed context, loaded into the OpenEvolve program database and mutation prompt. \\
Candidate contract & Candidate must expose \texttt{run\_attack} and return an image that can be evaluated externally. & Same contract; malformed or failing candidates receive penalty metrics. \\
Objective interface & Uses the EvoHash evaluation package directly with lower $S$ treated as better. & Exposes \texttt{combined\_score=-S} because OpenEvolve maximizes its framework-native score. \\
Search structure & Stores lineage and evaluation outputs for generated candidates. & Uses MAP-Elites-style archive, islands, and behavior features based on ASR, $L_2$, and queries. \\
Final artifact & A Python attack program scored by the external evaluator. & A Python attack program scored by the external evaluator. \\
\bottomrule
\end{tabularx}
\end{table}

\textbf{Seed Algorithm Portfolio.} The seed portfolio contains five attacks. SimBA tests signed basis directions greedily \cite{guo2019simpleblackboxadversarialattacks}. NES estimates a gradient from random perturbations \cite{ilyas2018blackboxadversarialattackslimited}. ZO-SignSGD is a local zeroth-order sign-gradient baseline related to query-efficient sign-gradient black-box attacks \cite{liu2018signsgd}. The Prokos seed is an adaptation of the targeted perceptual-hash attack family introduced by Prokos et al. \cite{285409}. ATKScopes uses multiresolution perturbations optimized with Adam \cite{kingma2017adammethodstochasticoptimization} and is the strongest scale-aware seed in the portfolio \cite{307842}. Each seed attack is evaluated in an original configuration and in a per-hash optimized configuration.

\section{Dataset}
\label{sec:dataset}

All image pairs are sampled from ImageNet \cite{5206848}. The benchmark contains 30 pairs and is used for hyperparameter optimization, program evolution, and reported comparisons. All images are resized to $256\times256$ using bilinear interpolation and represented in RGB with pixel values in $[0,1]$ when computing $L_2$. The reported tables correspond to the public benchmark split used by the evaluation artifacts. Because the same 30 pairs are used for search and reporting, the results should be interpreted as controlled benchmark evidence rather than a deployment-scale generalization claim.

\section{Experiments}
\label{sec:experiments}

\subsection{Stage 1: ASR Maximization via Hyperparameter Optimization}

The first experimental stage evaluates the five hand-written seed attacks for each hash under the hash-specific query budget. This stage prioritizes reliable threshold reaching, as measured by Eq.~\ref{eq:asr}, and provides the original seed results summarized in Table~\ref{tab:main_stage2}.

\subsection{Stage 2: Joint Quality--Efficiency Optimization via Composite Score}

The second stage tunes attack configurations independently for each hash and ranks viable candidates by the composite score in Eq.~\ref{eq:score}. The score combines pixel-space $L_2$, query count, and ASR, while LPIPS is retained as a separate reporting metric. The optimized configurations form the seed programs for the evolution stage.

\subsection{Stage 3: LLM-Driven Evolutionary Search}

The third stage uses GigaEvolve and OpenEvolve to mutate Python attack programs rather than only scalar parameters. Every generated candidate is evaluated by the same external evaluator on the 30-pair benchmark. Invalid programs, runtime failures, malformed results, and zero-success candidates receive evaluator penalties rather than manually curated scores.

\noindent\textbf{Evolution search configuration.}
Program evolution is budgeted to approximately $1{,}000$ candidate programs per hash per backend. OpenEvolve runs $1{,}000$ iterations with one candidate per iteration; GigaEvolve runs $100$ iterations with up to $10$ candidates per iteration. Both backends use the same mutation model, Qwen3.6-27B served in FP8. Each candidate is evaluated on the same 30-pair benchmark under the hash-specific query budget, so the numbers in Tables~\ref{tab:main_evolved} and~\ref{tab:best_vs_evolved} correspond to a single evolution run with a single random seed per hash per backend; cross-seed variance is not reported in this study. The seed-configuration stage (Stage~2) and the evolution stage were not run under a matched total-evaluation budget, so the comparison against optimized seeds is not compute-matched; a compute-matched seed search is left to future work.

\subsection{Baselines}

\subsubsection{Single-Stage Soft-Label Attacks}

The single-stage soft-label baselines are SimBA, NES, and ZO-SignSGD. They use oracle distance feedback throughout the attack and are evaluated in original and optimized configurations when available.

\subsubsection{PHA Attack Systems}

We use Prokos and ATKScopes. Prokos represents prior targeted hash-attack methodology, while ATKScopes supplies the strongest scale-aware seed in this benchmark.

The standard query budgets are 10,000 queries per pair for pHash and PDQ, 20,000 for PhotoDNA, and 30,000 for NeuralHash. LPIPS is computed only for reporting, not for search. NeuralHash receives asymmetric treatment for two reasons. First, in preliminary original-configuration runs on NeuralHash only ATKScopes reached a non-trivial success rate (ASR $0.50$); the other four seeds did not exceed ASR $0.13$ (NES $0.00$, Prokos $0.03$, SimBA $0.13$, ZO-SignSGD $0.03$), so tuning them offered little prospect of a viable configuration. Second, NeuralHash is substantially more expensive per query in the current implementation, which makes full hyperparameter optimization of low-ASR seeds not cost-effective. We therefore report full configuration optimization for NeuralHash only for ATKScopes and retain the other seeds as original baselines.

\begin{table}[t]
\centering
\caption{Optimization and evolution protocol.}
\label{tab:experiment_protocol}
\footnotesize
\begin{tabularx}{\textwidth}{@{}lY@{}}
\toprule
Stage & Protocol \\
\midrule
Original seed evaluation & Run five hand-written black-box seed attacks for each hash whenever the implementation is available under the hash-specific budget. \\
Configuration optimization & Tune attack hyperparameters independently for each hash and rank candidates by $S$ after first requiring reliable threshold-reaching behavior. \\
Program evolution & Use optimized attack programs as seeds for GigaEvolve and OpenEvolve; evaluate generated candidates on the 30-pair benchmark. \\
Metric recomputation & Recompute ASR, $L_2$, LPIPS, queries, final hash distance, and $S$ from returned images with the same external evaluator. \\
\bottomrule
\end{tabularx}
\end{table}

\section{Results}
\label{sec:results}

\subsection{Main Results}

Table~\ref{tab:main_stage2} summarizes the best original and optimized seed configurations. Table~\ref{tab:main_evolved} reports the evolved candidates, and Table~\ref{tab:best_vs_evolved} compares the best evolved program with the best optimized seed.

The results in this section are in-sample benchmark-selection results: the same 30 pairs are used for tuning, evolution, and reporting. ASR values should therefore be read as small-sample counts. ASR values of 1.00, 0.83, 0.70, and 0.67 correspond to 30/30, 25/30, 21/30, and 20/30 successes; Wilson 95\% intervals are [0.89, 1.00], [0.66, 0.93], [0.52, 0.83], and [0.49, 0.81], respectively.

\begin{table}[!htbp]
\centering
\caption{Best seed attacks before and after configuration optimization. Lower score is better; $\Delta S$ is computed between the best original score and the best optimized score in the row.}
\label{tab:main_stage2}
\small
\setlength{\tabcolsep}{3pt}
\begin{tabular*}{\textwidth}{@{\extracolsep{\fill}}llccclcc@{}}
\toprule
Hash & Best original & $S_\mathrm{orig}$ & ASR & Best optimized & $S_\mathrm{opt}$ & ASR & $\Delta S$ \\
\midrule
pHash & ATKScopes & 111.3 & 0.90 & ATKScopes & 86.1 & 1.00 & 22.7\% \\
PDQ & Prokos & 87.6 & 1.00 & ATKScopes & 44.2 & 1.00 & 49.5\% \\
PhotoDNA & SimBA & 163.8 & 1.00 & SimBA & 153.6 & 1.00 & 6.2\% \\
NeuralHash & ATKScopes & 1021.8 & 0.50 & ATKScopes & 654.9 & 0.70 & 35.9\% \\
\bottomrule
\end{tabular*}
\end{table}

\begin{table}[!htbp]
\centering
\caption{Evolved attack candidates.}
\label{tab:main_evolved}
\small
\begin{tabular*}{\textwidth}{@{\extracolsep{\fill}}llccccc@{}}
\toprule
Hash & Candidate & ASR & $L_2$ & LPIPS & Queries & $S$ \\
\midrule
pHash & OpenEvolve & 1.00 & 27.38 & 0.150 & 1471 & 56.80 \\
pHash & GigaEvolve & 1.00 & 38.94 & 0.082 & 910 & 57.13 \\
PDQ & OpenEvolve & 1.00 & 13.80 & 0.111 & 676 & 27.31 \\
PDQ & GigaEvolve & 1.00 & 25.48 & 0.101 & 628 & 38.04 \\
PhotoDNA & GigaEvolve & 1.00 & 104.50 & 0.572 & 1836 & 141.22 \\
PhotoDNA & OpenEvolve & 1.00 & 100.45 & 0.596 & 2297 & 146.40 \\
NeuralHash & GigaEvolve & 0.83 & 133.07 & 0.697 & 9382 & 384.84 \\
NeuralHash & OpenEvolve & 0.67 & 127.22 & 0.682 & 7396 & 412.72 \\
\bottomrule
\end{tabular*}
\end{table}

\begin{table}[!htbp]
\centering
\caption{Best optimized seed versus best evolved attack. Improvement is score reduction.}
\label{tab:best_vs_evolved}
\small
\begin{tabular*}{\textwidth}{@{\extracolsep{\fill}}llclcc@{}}
\toprule
Hash & Best optimized seed & $S_\mathrm{opt}$ & Best evolved & $S_\mathrm{evo}$ & Improvement \\
\midrule
pHash & ATKScopes & 86.08 & OpenEvolve & 56.80 & 34.0\% \\
PDQ & ATKScopes & 44.25 & OpenEvolve & 27.31 & 38.3\% \\
PhotoDNA & SimBA & 153.64 & GigaEvolve & 141.22 & 8.1\% \\
NeuralHash & ATKScopes & 654.89 & GigaEvolve & 384.84 & 41.2\% \\
\bottomrule
\end{tabular*}
\end{table}

The main conclusion is not that evolution always improves every individual metric. The improvement is in the composite score. Relative to the best optimized seed, the best-by-score evolved program improves all reported raw axes for PDQ and NeuralHash, while pHash and PhotoDNA trade more queries for lower visual cost. For NeuralHash, only ATKScopes receives full seed-configuration optimization (the other four seeds did not exceed ASR $0.13$ in their original configurations, as noted in Section~\ref{sec:experiments}), so the 41.2\% reduction is against the optimized ATKScopes seed rather than a fully optimized five-seed portfolio; the evolved ASR remains 25/30. For PhotoDNA and NeuralHash, LPIPS remains high, so these attacks should be interpreted as benchmark collisions rather than visually imperceptible examples.

\subsection{Program-Level Changes}
\label{sec:evolved_changes}

The evolved programs were not mere parameter variants. Table~\ref{tab:evolved_changes} summarizes the main structural changes found in the final candidates. The recurring pattern is hash-specific search-space adaptation followed by a refinement phase that reduces distortion after reaching the collision threshold.

\begin{table}[!htbp]
\centering
\caption{Qualitative changes introduced by evolved attack programs.}
\label{tab:evolved_changes}
\footnotesize
\setlength{\tabcolsep}{2pt}
\begin{tabularx}{\textwidth}{@{}lllY@{}}
\toprule
Hash & Candidate & Ancestor & Main program-level changes \\
\midrule
pHash & OpenEvolve & ATKScopes & Collision post-processing; smaller-$L_2$ selection; DCT coefficient shrinkage; luminance-consistent projection \\
pHash & GigaEvolve & ATKScopes & Sequential low-frequency DCT traversal; decaying step size; gradient clipping; final $L_2$ refinement \\
PDQ & OpenEvolve & ATKScopes & Luminance-oriented search; coordinate importance estimates; DCT delta scaling; small-coefficient pruning \\
PDQ & GigaEvolve & ATKScopes & Single grayscale DCT variable shared across channels; normalized RGB perturbation scale \\
PhotoDNA & GigaEvolve & ATKScopes & Central-crop PhotoDNA search; patch-DCT coordinates; plateau handling while preserving optimizer state \\
PhotoDNA & OpenEvolve & SimBA & Adaptive SimBA step; local refinement over successful coordinates; greedy DCT-component reduction \\
NeuralHash & GigaEvolve & ATKScopes & Multi-coordinate DCT probes; adaptive step expansion/shrinkage; binary search back to source after success \\
NeuralHash & OpenEvolve & ATKScopes & Probing step; multi-coordinate restarts near threshold; accumulated successful DCT perturbations \\
\bottomrule
\end{tabularx}
\end{table}

\begin{figure}[!htbp]
\centering
\includegraphics[width=0.8\textwidth]{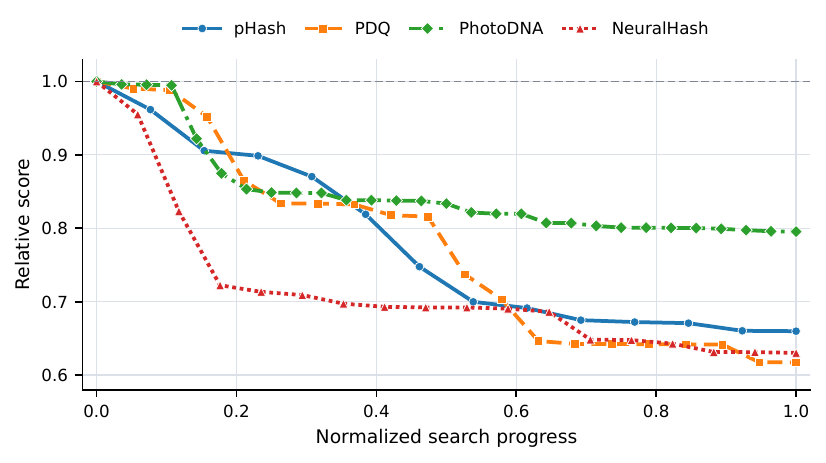}
\caption{Normalized OpenEvolve champion score during program evolution. Lower values indicate improvement over the initial seed program.}
\label{fig:openevolve_score_progress}
\end{figure}

Figure~\ref{fig:openevolve_score_progress} illustrates the OpenEvolve search trajectory. The plotted value is the best score observed so far along the champion chain, normalized by the score of the initial seed program for the corresponding hash. Values below 1.0 indicate a lower-score candidate during program evolution.

\section{Discussion}
\label{sec:discussion}

\noindent\textbf{Program evolution improves beyond seed tuning.}
Configuration optimization substantially improves pHash, PDQ, and NeuralHash, but evolved programs still reduce the final score for every hash. This indicates that the useful search space includes algorithmic changes rather than only scalar hyperparameters.

\noindent\textbf{The improvement is score-level, not metric-wise monotonic.}
The evolved candidates do not always reduce $L_2$, LPIPS, and query count simultaneously. For example, a candidate may spend more queries to obtain a substantially cleaner image, or may preserve ASR while reducing distortion. Reporting all metrics is therefore necessary. The composite score should be interpreted as a ranking criterion, not as a replacement for the individual measurements.

\noindent\textbf{Hash geometry matters.}
pHash and PDQ are easier targets in this benchmark once attacks operate in a suitable low-frequency search space. PhotoDNA requires larger perturbations, but evolved candidates still reduce score relative to optimized seeds. NeuralHash is the hardest target in this benchmark: its best evolved ASR is 0.83, and it requires substantially more queries than the classical hashes. This is consistent with the fact that the useful geometry is less directly aligned with simple DCT threshold comparisons.

\noindent\textbf{The two evolution backends are complementary.}
OpenEvolve gives the best score for pHash and PDQ, while the GigaEvo-based pipeline gives the best score for PhotoDNA and NeuralHash. This suggests that the benefit is not tied to a single evolution backend. The common driver is the ability to generate and evaluate program-level attack variants under a strict external metric.

\noindent\textbf{Defensive interpretation.}
The benchmark should be read as an audit protocol, not only as a catalog of attacks. A perceptual-hash system that appears robust against default attack settings may look different after per-hash hyperparameter tuning and program evolution. This means that threshold selection, preprocessing, and consistency checks should be evaluated against strengthened attacks rather than against natural transformations or fixed baselines alone. The present study does not claim deployment-ready defense parameters; such parameters require domain-specific false-positive and benign-recall evaluation at substantially larger scale.

\section{Conclusion}
\label{sec:conclusion}

We presented a unified black-box framework for targeted second-image attacks on perceptual image hashes and used it to evaluate fixed seed attacks, optimized configurations, and LLM-evolved attack programs. On the in-sample 30-pair benchmark, the best evolved candidates reduce the composite score relative to the best optimized seed by 34.0\% for pHash, 38.3\% for PDQ, 8.1\% for PhotoDNA, and 41.2\% for NeuralHash. The results support a narrower claim: under this controlled evaluator, searching over attack programs can find lower-score candidates than tuning fixed black-box attacks alone. Held-out evaluation, repeated evolution runs, and compute-matched seed tuning are needed before claiming generalization.

\section{Limitations}
\label{sec:limitations}

The study is empirical and depends on the available hash implementations and preprocessing. Industrial deployments may use different thresholds, rate limits, and decision logic. The benchmark contains 30 ImageNet pairs and the same split is used for search and reporting, so the results are controlled benchmark evidence rather than deployment-scale generalization. NeuralHash is computationally expensive in the current framework; only ATKScopes receives full configuration optimization for that hash, as the remaining seeds did not exceed ASR $0.13$ at baseline. The reported evolution results come from a single evolution run with a single random seed per hash per backend, so cross-seed variance and paired significance are not reported. The experiments cover targeted second-image collisions only, not untargeted evasion or hash inversion.

\section{Future Work}
\label{sec:future}

Future work should evaluate larger and more diverse image sets, including domains whose low-level statistics differ from ImageNet. A second direction is to extend the same evaluator to untargeted evasion objectives and to realistic API constraints such as rate limits, latency, hard-label feedback, and noisy feedback. A third direction is defensive evaluation: the evolved attacks can be used to test preprocessing, threshold selection, and consistency checks under a stronger benchmark than fixed default attacks. Finally, equipping deployed hash-matching decisions with calibrated uncertainty estimates or statistical guarantees, as studied in the uncertainty-quantification literature \cite{fishkov2025uncertaintyquantificationregressionusing}, is a promising direction toward the instance-wise guarantees motivated in the introduction.

\bibliographystyle{splncs04}
\bibliography{references}
\end{document}